\documentclass{article}

\usepackage{arxiv}

\usepackage{graphicx}                         
\usepackage{dcolumn}                          
\usepackage[version=3]{mhchem}                
\usepackage[T1]{fontenc}                      

\usepackage{epsfig}
\usepackage{amsmath}
\newcommand{\angstrom}{\mbox{\normalfont\AA}}
\usepackage{epstopdf}
\usepackage{amsbsy}
\usepackage{multirow}
\usepackage{bm}
\usepackage{caption}
\usepackage{siunitx}
\usepackage{CJK}
\usepackage{textcomp}
\usepackage{comment}
\usepackage{soul}
\usepackage[ruled,vlined]{algorithm2e}
\usepackage{caption}
\usepackage{subcaption}
\usepackage{xcolor}
\usepackage{array}
\usepackage{comment}
\usepackage{pdfpages}

\newcolumntype{P}[1]{>{\centering\arraybackslash}p{#1}}

\title{Understanding Creep in Vitrimers: Insights from Molecular Dynamics Simulations}

\author{ {Gurmeet Singh, Veera Sundararaghavan}\thanks{Corresponding author: Prof. Sundararaghavan, Email: veeras@umich.edu, Tel: 734-615-7242} \\
	Department of Aerospace Engineering\\
	University of Michigan\\
	Ann Arbor, MI \\
	\texttt{gmsingh@umich.edu}\\
	\texttt{veeras@umich.edu} \\
	\And
	{\hspace{1mm}Vikas Varshney} \\
	Materials and Manufacturing Directorate\\
	Air Force Research Laboratory\\
	Wright-Patterson Air Force Base, OH \\
	\texttt{vikas.varshney.2@us.af.mil} \\
}

\renewcommand{\shorttitle}{\textit{arXiv} Paper}

\begin{document}
\maketitle

\begin{abstract}
Vitrimers offer a promising sustainable alternative to conventional epoxies due to their recyclability. Vitrimers are covalent adaptive networks where some bonds can break and reform above the vitrimer transition temperature. While this can lead to desirable behavior such as malleability, this also leads to undesirable rheological behavior such as low-temperature creep. In this work, we investigate the molecular mechanisms of the creep of vitrimers using molecular dynamics simulations. The interplay between dynamic bonding with mechanical loading is modeled using a topology-based reaction scheme. The creep behavior is compared against cross-linked epoxies with dynamic reactions to understand the unique aspects related to dynamic bonding. It is found that the free volume that arises from tensile loads is reduced in vitrimers through dynamic bond rearrangement. An important feature that explains the difference in secondary creep behavior between conventional epoxies and vitrimers is the orientation of the dynamic bonds during loading. In vitrimers, the dynamic bonds preferentially align orthogonal to the loading axis, decreasing the axial stiffness during secondary creep, resulting in larger creep strain compared to epoxies. Over longer timescales, such increased strain leads to void growth, resulting in tertiary creep. Thus, chemistry changes or additives that can prevent the initial realignment of dynamic bonds, and therefore subsequent void growth, can be an effective strategy to mitigate creep in vitrimers. 
\end{abstract}


\section{Introduction}

Epoxy is a thermoset polymer that is widely used in automobile, aerospace, robotics, and wind energy industries\cite{zabihi2018technical,saba2016recent}. Due to their thermal and chemical stability, they have played an important role in the emergence of advanced high-performance composites. However, the thermoset nature of epoxies has limited their life cycle due to the lack of damage mitigation or recycling capabilities\cite{Arias2003,singh2017damage}. Another class of polymers, thermoplastics, are easy to recycle but are limited by their thermomechanical performance in critical structural applications\cite{penumakala2020critical,trivedi2021graphene}. Vitrimers are a new class of polymers that offer the best of both thermosets and thermoplastics via dynamic cross-link reactions in the polymer network. They behave like a cross--linked thermoset at room temperatures and demonstrate malleable properties of a thermoplastic when heated beyond a temperature where dynamic bonds become active\cite{vitrimer,capelot2012catalytic,Zheng2021}. This ability of vitrimers makes them a promising candidate towards damage mitigation during operation and recyclability afterwards\cite{zhang2016advances,islam2021progress}. 

A major challenge associated with the usage of vitrimers is associated with low-temperature creep\cite{terryn2021review,li2022facile,hubbard2021vitrimer}. Creep is the deformation of the material with time under the application of constant stress. Creep strain in polymers is simplified to Findley's power law, $\epsilon = \epsilon_0 + \epsilon^+ t^n$, where $t$ is time and $\epsilon_0$, $\epsilon^+$ and $n$ are constants for a given stress level\cite{findley2013creep}. Polymers, whose glass transition (glassy to rubbery transition) temperatures are relatively closer to room temperature, are prone to  experiencing creep at room temperature. Often, creep can lead to an undesirable amount of deformations that compromise the integrity and function of a structure\cite{bradley1997viscoelastic,sa2011creep,brinson2008polymer}. Therefore, it is important to understand the creep behavior and underlying mechanisms for improved molecular design of vitrimers.

Creep in polymers and fiber composites is a well-studied phenomenon both experimentally\cite{bradley1997viscoelastic,sa2011creep,brinson2008polymer,bradley1998viscoelastic} and computationally\cite{Riggleman2008,Jian2019,Chang2021,Plaseied2008}.  Creep under uniaxial tension is found to follow three stages. In primary creep, the strain increases at a rapid rate during initial loading but continues to slow down over time. The second stage, termed secondary creep, is characterized as a region of uniform strain rate. Tertiary creep is the final stage of creep where the material strain accelerates and leads to a rupture. Molecular scale experiments by Lee et al. on crosslinked poly(methyl methacrylate) reinforce the notion that  stress-induced chain mobility allows polymer glasses to flow during creep\cite{lee2009direct}. In addition, Bradley et al. found that the creep in vinylester resins reduces with the  duration of resin curing due to higher cross-linking\cite{bradley1998viscoelastic}. While adding reinforcing fibers was found to reduce creep, the exponent $n$ was found to be largely unchanged. 

Unlike conventional epoxies, understanding of the creep in vitrimers is rather limited. It is difficult to probe and investigate the underlying dynamic bonding mechanisms in vitrimers experimentally and this is where molecular simulations provide an exciting alternative to further such understandings\cite{Sun2020,Park2020}. A recent experimental study by Hubbard et al. sheds light on the possible molecular mechanisms in the stages of vitrimer creep, where it is postulated that secondary creep in vitrimers is associated with network rearrangement due to dynamic reactions\cite{hubbard2022creep}. It was also noted that at low temperatures and catalyst concentrations, vitrimers simply behave as a traditional epoxy material. Dynamic cross-linking reactions tend to accelerate beyond the topology freezing temperature ($T_V$), however, a small extent of these reactions at lower temperatures can influence creep behavior\cite{hubbard2021vitrimer,liu2017catalyst}. 

In this work, we employ all--atom molecular dynamics (MD) simulations to study the creep behavior in vitrimers, which is carried out for the first time to the best of the authors' knowledge.  The MD framework has been widely utilized to predict the properties of metals\cite{Jiao2015,singh2021understanding} and polymers including mechanical\cite{BANDYOPADHYAY20112445}, thermal expansion\cite{msslepoxy3}, thermal conductivity\cite{msslepoxy2,VARSHNEY20093378}, heat capacity\cite{VARSHNEY20093378}, and glass transition\cite{BANDYOPADHYAY20112445,singh2020modeling} properties. In particular, MD simulations have also been utilized to model the creep behavior of metals\cite{Jiao2015,Keblinski1998,Yamakov2002,simoes2006molecular} as well as polymers\cite{Riggleman2008,Chang2021,riggleman2007free,sahputra2015creep}. 
Several of these molecular models, specifically polymers, simplify atomic interactions using coarse-grained/bead-spring models to model chain dynamics. While such models capture conformational changes, they poorly describe chain-to-chain interactions that determine the free volume evolution. A recent all-atom MD study of creep shows that secondary creep is mechanistically related to void nucleation, while tertiary creep is related to void growth and coalescence\cite{Bowman2019}. Furthermore, MD simulations by Li et al. describe creep in epoxies in the context of the free-volume change theory of Fox and Flory. With increasing stress and temperature, the creep correlated with increases in the free volume in the simulation cell\cite{li2021uniaxial}. It is to be noted that the time scale of creep at the macro scale is in the range of hours or even days. However, for the purpose of understanding the underlying deformation mechanism of creep, MD simulations of creep need to be performed using elevated stress and/or temperature, and at high strain rates.

The primary challenge for the MD approach is the modeling of the temperature-dependent reversible cross--link reactions. Exchange reactions have been modeled in the past via methods such as embedding Monte Carlo (MC) moves into molecular dynamics,  fully MD (using specialized reactive potentials), or fully MC simulations to simulate bond swaps\cite{ciarella2018dynamics,Oyarzun2018,wu2019dynamics,Smallenburg2013,Sciortino2017}. These simulations have typically employed coarse-grained (bead--spring) models that provide high computational efficiency while approximating the mechanical response. For more quantitative modeling, all--atom MD methods are attractive\cite{Sun2020,Yang2016}. Previously, bond exchange reactions in all--atom MD have been implemented using distance--based reaction schemes based on pre and post--reaction templates\cite{singh2020modeling,gissinger2017modeling}. The approach accelerates the slow reaction dynamics and allows the modeling of mechanical property changes in vitrimers during thermal cycling. 

Typically, creep in epoxies is simulated by applying stress and letting the system evolve over time under NPT dynamics\cite{li2021uniaxial}. However, in a vitrimer system with dynamic bond exchange reactions, NPT dynamics can become unstable due to energy fluctuations caused by local topology modification whenever dynamic cross-linking reactions occur. In this work, we devise a new strategy to simulate the creep behavior of vitrimer with dynamic reactions by alternating loading and equilibration steps. In parallel to the experimental work by Hubbard et al., the approach is used to differentiate creep response in vitrimers and traditional epoxies to identify mechanisms fundamentally derived from their dynamic bond exchange reactions\cite{vitrimer,capelot2012catalytic,hubbard2022creep,denissen2016vitrimers}.

\section{Methods}
We employ classical MD simulations to simulate the creep behavior of vitrimer system, composed of diglycidyl ether of bisphenol-A (DGEBA) cross-linked with 4-aminophenyl disulfide (AFD), the chemical structures shown in Figure 1(a). Large-scale atomic/molecular massively parallel simulator  (LAMMPS)\cite{plimpton1995fast,thompson2022lammps} is used to carry out all of the simulations in this study. consistent valence force field (CVFF) is assigned to all the atoms with energy contributions from the pair, bond, angle, dihedral, and improper interaction terms\cite{dauber1988structure}. Additionally, the energy contributions from  the non-bonded interactions are modeled using Lennard-Jones (LJ) and Coulombic pairwise interaction with a cutoff of $12~\angstrom$. A time step of 1 fs is used for all of the MD simulations in this work.

\begin{figure}[h]
    \centering
    \includegraphics[trim={0.25cm 0cm 0cm 8.0cm},clip=true ,width=1.0\textwidth]{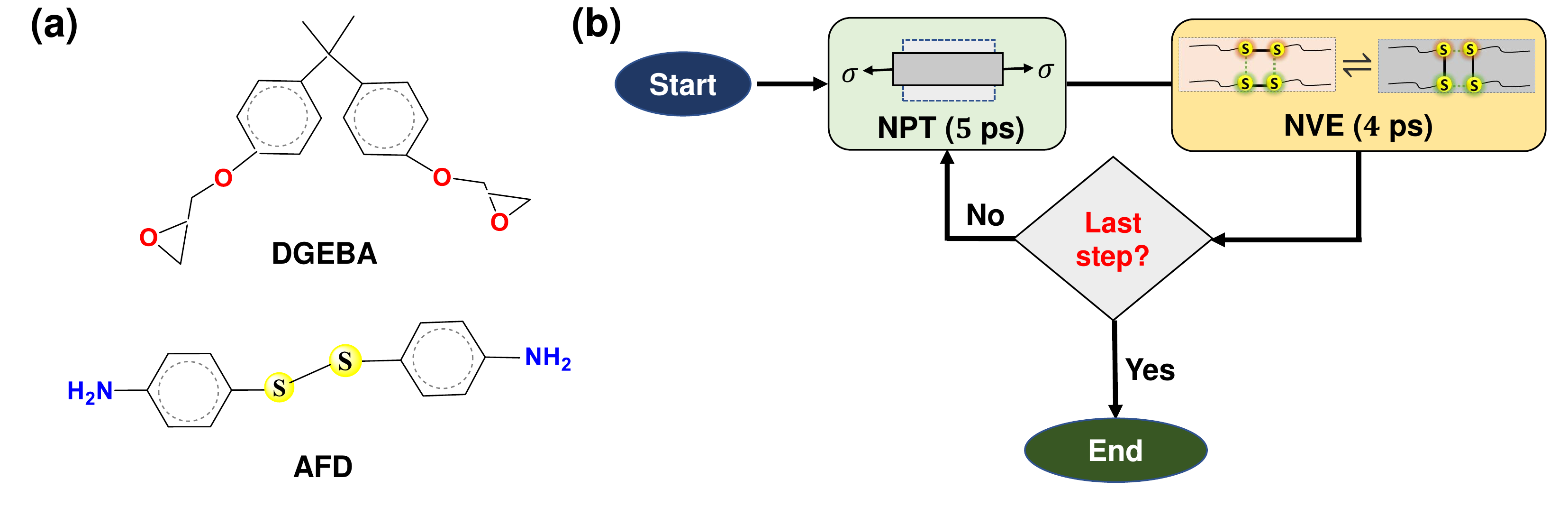}
    \caption{(a) Monomer structures, and (b) Methodology for simulating creep with dynamic S$-$S reactions}
    \label{fig:methods}
\end{figure}

\subsection{Polymer system preparation}
The polymer system for MD simulations is prepared using polymerization reactions (primary and secondary amine reactions) for curing the monomer mixture. In this approach, the \emph{fix bond/react} feature in LAMMPS is utilized that enables the modeling of reactions by changing the local topology\cite{gissinger2017modeling,gissinger2020reacter}. First, a mixture of monomers with two DGEBA units and one AFD unit is constructed. The typical synthetic epoxy to hardener stoichiometric ratio of 2:1 is employed\cite{singh2017damage,varshney2008molecular}. The repeating unit that contains the monomer mixture is shown in Figure S.1 of supporting information. We repeat this unit by $8\times8\times8$ to get a simulation box with 1,024 DGEBA and 512 AFD monomer units, with a total of 68,608 atoms. Periodic boundary conditions are applied in all three directions. Then, the constructed monomer mixture is equilibrated using the NVT ensemble (constant number of particles, volume, and temperature). Subsequently, the mixture is equilibrated to a density of $1.0~ \text{gcm}^{-3}$ using an NPT ensemble (constant pressure and temperature). Nose-Hoover thermostat and barostat are used to maintain the temperature and pressure in the simulation box, respectively. The pre-- and  post--reaction templates are prepared for both primary and secondary amine reactions and can be found in our previous work\cite{singh2020modeling}. The cutoffs between N and C atoms are set to 3.5 \si{\angstrom} and 5.0 \si{\angstrom} for initiating the primary and secondary amine reactions, respectively. The mixture is then allowed to have primary and secondary amine reactions under the NVT ensemble. After the system is cured to $95\%$, it results in a system with internal stresses in the box. Therefore, to relieve these stresses, the system is annealed by undertaking the box to cooling and heating cycles of 1 K and 600 K, respectively with a simulation time of 50 ps at each of these temperatures under NPT (pressure of 1 bar), as also reported previously\cite{singh2020modeling}. This process releases the stresses built in the box and brings the system to an equilibrated density of $\rho=1.184~\text{gcm}^{-3}$ at 1 K, and the converged density at 300 K of $\rho=1.159~\text{gcm}^{-3}$ which is in agreement with values for a typical epoxy system from literature\cite{Afzal2021}. The resulting MD simulation box is shown in Figure\ref{fig:rxnNorxn_strain}(a), where the inset shows a section of the chain segment highlighting the presence of singly and doubly reacted Nitrogen atoms as a result of curing. The inset also shows a potential site for a disulfide bond exchange reaction where two S$-$S bond chain segments come in close vicinity to result in an exchange of chains as illustrated in Figure\ref{fig:rxnNorxn_strain}(b). These are the characteristic reactions of this vitrimer system which accelerate above its topology transition temperature and result in the rearrangement of the network. The disulfide reactions are modeled using the topology-based update, and the pre- and post-reaction templates for these vitrimer reactions are shown in Figure S.2 of the supplementary information. The reaction occurs when any two sulfur atoms from two disulfide pairs come within a cutoff of 4.12 \si{\angstrom}. Then the reaction can proceed with an assigned probability, where, the probability of 0.0 implies no S--S reactions while 1.0 means all such eligible disulfide pairs can have a bond exchange reaction.

\subsection{Simulating creep}
Computational experiments of creep are simulated under constant stress applied in the loading axis while other transverse directions are kept stress-free\cite{Li2021,Tam2021,simoes2006molecular}. In this work, this is achieved using Nose-Hoover anisotropic barostat and thermostat under NPT conditions on the box. An anisotropic barostat allows individual stress components to be prescribed on the box. For simulating uniaxial creep, the prescribed state of true stress is: $\sigma_{yy}=\sigma_o,~\sigma_{xx}=\sigma_{zz}=0$ where the constant true stress $\sigma_o$ is maintained in $y-$axis. All three shear stress components are left unspecified which implies the shear strains on the box are kept to zero, therefore, the box stays orthogonal during the creep simulations. 

For an epoxy system, creep could be simulated by applying a constant value of stress and letting the system evolve over time under NPT dynamics. However, in a vitrimer system with dynamic bond exchange reactions, NPT dynamics can become unstable under applied pressure due to local fluctuations in temperature and pressure tensor caused by local molecular topology changes during reaction events\cite{gissinger2017modeling,gissinger2020reacter}. Furthermore, under stress-controlled loading (such as in creep), the box can change the dimensions suddenly and the dynamic cross-linking reactions can result in the loss of bonds or atom images across box boundaries which can lead to simulations failure. In the case of strain-controlled loading, given it is small, the dynamic bond reactions can take place while deforming the box slowly as reported in our previous work\cite{singh2020modeling}. In the present study, to simulate the creep behavior of vitrimers, which is a stress-controlled loading, we have devised a new strategy as shown in Figure \ref{fig:methods}(b). Here, we break the application of stress into two parts, first, an NPT is run for 5 ps followed by an NVE for 4 ps under which the dynamic S$-$S reactions happen. The S--S reactions are allowed to happen only once in this reaction step, and which is controlled by specifying the reaction frequency in \emph{fix bond/react} card in LAMMPS. The NVE step allows the system to have reactions when there is no deformation happening (constant volume) and adequate time to relax the system after the reactions occur. The NPT step is then invoked where the system deforms under creep, and these steps are repeated. For comparison with the case of epoxy, the effect of the additional NVE step on creep response is discussed in Figure S.3 in supporting information. To accelerate the creep phenomenon in MD simulations, we apply a high value of uniaxial stress at high temperatures. The value of the applied stress dictates the resulting creep response of vitrimers, the role of which is also discussed in the later sections. The resulting response is characterized by the stretch ratio ($\lambda$) in the loading direction, which is defined as:
\begin{equation}
    \lambda = \frac{l}{l_o}
\end{equation}

where, $l$ is the current length and $l_o$ is the initial or undeformed length of the simulation box along the loading direction. In this work, the following studies are conducted: (a) the comparison of the creep response of a vitrimer and an epoxy; (b) the influence of applied stress or loading on vitrimer creep; and (c) the effect of S--S reaction probability on the creep response of vitrimers.

\subsection{Free volume of voids}
The free volume or the void volume is the region of space where atoms are not present and it is computed using an \emph{alpha-shape} method\cite{edelsbrunner1994three} implemented in OVITO software\cite{stukowski2014computational}. All the simulations are visualized using OVITO by performing surface mesh construction analysis. The probing sphere has a finite radius and for the analysis in the present study, we have considered it to be 3.5 \si{\angstrom} and the details of which can be found from OVITO documentation\cite{stukowski2014computational}. This approach helps us identify the actual material volume as well as the void volume for any time instance during the simulation of creep. The volume fraction of the void is computed using a Python script for OVITO.  The percent volume fraction of the void ($V_f^{void}$) is defined as:
\begin{equation}
    V_f^{void}=\frac{V_{void}}{V_{cell}}\times 100
\end{equation}
where, $V_{void}$ and $V_{cell}$ are the volume of the empty region and the total volume of the MD simulation cell at a given time instance, respectively. 
\subsection{Bond orientation}
The vitrimer undergoes large deformation under creep in these simulations, therefore, it is important to probe the molecular mechanisms contributing to different aspects of the creep deformation and system evolution. MD simulations offer the capabilities to look into various quantities during the system evolution. One such quantity of interest is the orientation of the disulfide bonds when they undergo dynamic cross-linking reactions during deformation. Bond orientations of the S$-$S bonds are computed for the entire simulation box and used to analyze the results of the vitrimer with both, dynamic reactions and with  reactions turned off as a baseline. The bond vector is computed using the coordinates of the two sulfur atoms:
\begin{equation}
    \bm{v}^\text{S-S} = \bm{x}_1-\bm{x}_2 
\end{equation}
Since the simulation box is periodic in all dimensions, the coordinates of the two sulfur atoms are corrected based on the image flag of the atom. The bond vector projection of an S$-$S bond pair on the loading axis ($y-axis$ in this case) is computed as: 
\begin{equation}\label{b_Vector}
    v^\text{S-S}_y = \Big|\bm{v}^\text{S-S}\cdot \bm{y}\Big|
\end{equation}
The average value of the  bond vector projection on the loading axis is evaluated as:
\begin{equation}\label{meanb_Vector}
    \overline{v}^\text{S-S}_y = \frac{1}{N} \sum _{i=1}^{N}\Big|\bm{v}_i^\text{S-S}\cdot \bm{y}\Big|
\end{equation}
where, $<\cdot>$ denotes the dot product of  two vectors, $\bm{y}=[0 ~1 ~0]^T$ is the loading axis vector and $N$ is the total number of S$-$S bonds in the simulation box. This computation is implemented in OVITO using the python scripting interface. First, all the S$-$S (disulfide) bonds are selected and then their connecting atoms are identified. The coordinates of the connecting sulfur atoms are used to evaluate the S$-$S bond orientation vector. 

\section{Results and Discussion}

Creep is an inherently slow phenomenon, beyond the time--scales of MD simulations when simulated at ambient conditions. Simulations of creep using atomistic simulations make an inherent assumption that the mechanisms that drive creep under ambient conditions also exist at extreme conditions at which accelerated creep occurs. Accelerated creep occurs at extreme conditions such as at high tensile loads close to or exceeding the yield stress and at higher temperatures where vitrimer reactions occur rapidly. In order to test all regimes of vitrimer creep, we employ extreme conditions in our simulations such that creep can occur rapidly and the mechanisms can be systematically analyzed. Another important assumption is that bond rupture of crosslinked epoxy bonds is not simulated. While such bond rupture occurs at high strains during extreme loading, it is necessary to suppress such mechanism of failure to study the actual mechanisms that might occur in ambient conditions where creep processes are much slower. In later subsections, we systematically reduce the extremity of loading to study the trends in the creep behavior as we move toward ambient conditions. 

\begin{figure}[h]
    \centering
    \includegraphics[trim={0.cm 2.1cm 0.cm 0.0cm},clip=true ,width=1.0\textwidth]{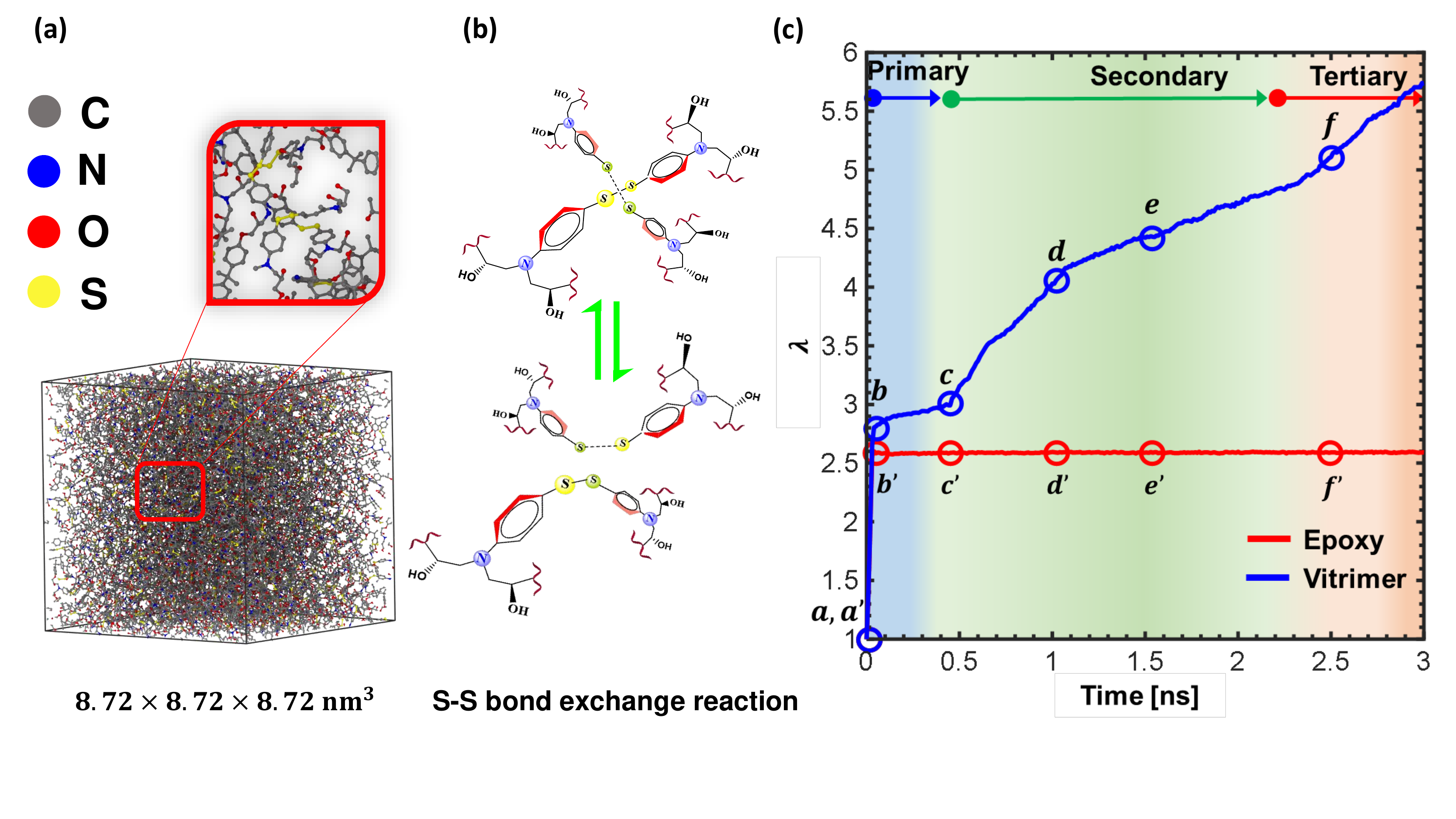}
    \caption{(a) Cured and annealed periodic polymer system shown with H atoms hidden, (b) schematic of a disulfide bond exchange reaction in the vitrimer system, and (c) comparison of stretch ratio ($\lambda$) response vs. time for vitrimer and epoxy under constant uniaxial stress of $500$ MPa at $600$ K, the points from $a(a')$ to $f(f')$ refer to key chosen snapshots during creep deformation as discussed in the text. Three creep regimes of a vitrimer are marked with different colors.}
    \label{fig:rxnNorxn_strain}
\end{figure}

\subsection{Creep in vitrimer vs. epoxy}

Figure \ref{fig:rxnNorxn_strain}(c) shows the time-dependent deformation response of the vitrimer with and without the dynamic bond exchange reactions under uniaxial stress of $500$ MPa at $600$ K. This is an extreme case where the temperatures are well above the topological transition temperature, hence, we specify the probability of the S$-$S bond reactions to be $p=1.0$. When the dynamic reactions are switched off, the material behaves like a conventional epoxy. Henceforth, the case with no reactions is thus referred to as the `epoxy' and the case with reactions is referred to as the `vitrimer'. With the application of tensile stress, both vitrimer and epoxy show an immediate increase in the strain, referred to as the primary creep response. The vitrimer is slightly more compliant than epoxy in this regime. The interesting difference arises after the primary creep wherein the epoxy does not show a significant increase in the creep strain. However, the vitrimer shows a marked increase in the strain after the primary regime. We refer to this regime as the secondary creep, which occurs in vitrimers due to dynamic reactions. A third regime called `tertiary creep' indicated as `$f$' in Figure \ref{fig:rxnNorxn_strain}(c) is also observed at which void growth is observed to occur (shown and discussed in detail later in the manuscript). These three regimes of creep are shown using different colors which a gradual transition between them.

\begin{figure}[h]
    \centering
    \centering
    \includegraphics[trim={0cm 0.0cm 8.cm 9cm},clip=true ,width=0.9\textwidth]{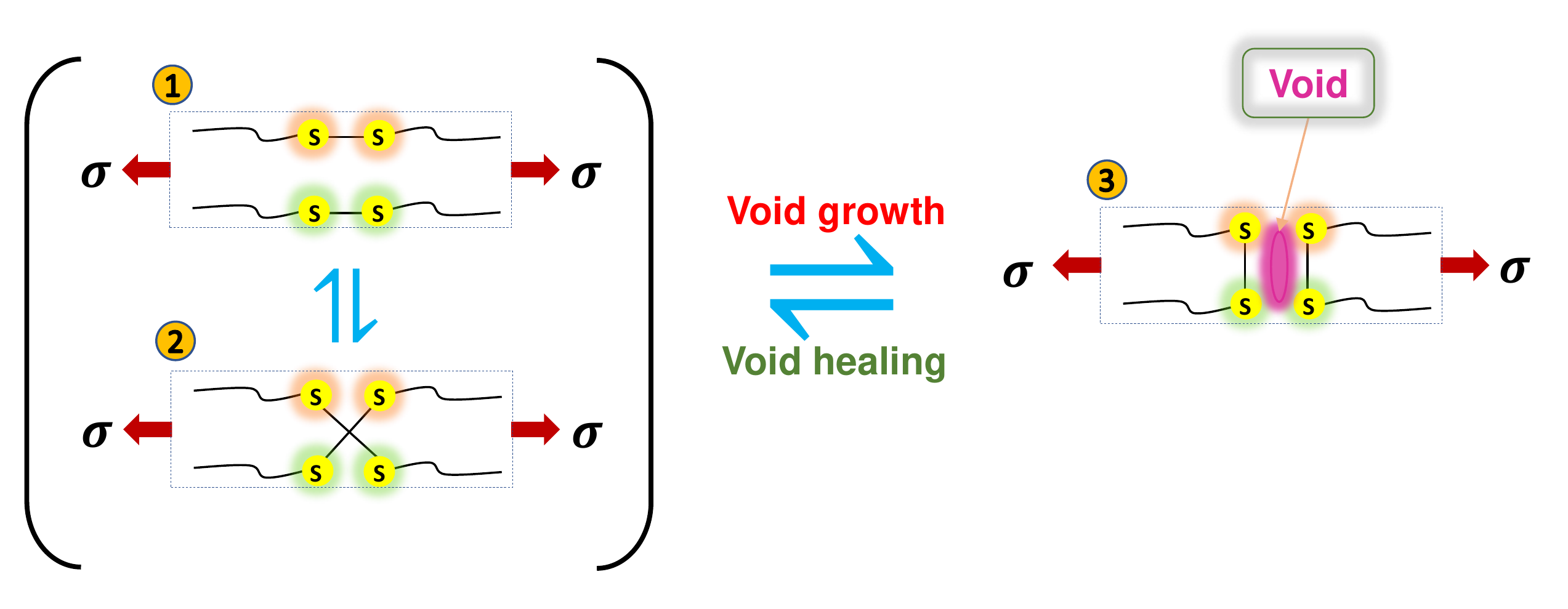}
    \caption{Possibilities of reaction pathways of the idealized chains opening due to S$-$S reactions under creep in vitrimers: \textcircled{1}-parallel, \textcircled{2}-crossed, and \textcircled{3}-perpendicular to the loading axis}
    \label{fig:S-S_schematic}
\end{figure}

For now, we focus on the mechanistic aspects of the secondary creep effect due to dynamic reactions in a vitrimer. Here, the types of chain arrangements are idealized into three configurations based on the orientation of the S$-$S bonds pairs in the vicinity (< 4.12 \si{\angstrom}) of each other, where the exchange between these configurations could lead to different outcomes on the bulk behavior of the vitrimer. To simplify the chain arrangements, we assume that the S$-$S bonds in the two chains  are aligned in parallel, perpendicular, or in a crossed formation with respect to the loading axis. Figure \ref{fig:S-S_schematic} shows the possible chain rearrangements due to S$-$S reactions between these three configurations. The dynamic exchange between configurations \textcircled{1} and \textcircled{2} leads to neither void healing nor growth since the axial loading can be accommodated by the chains in both of these configurations. 

On the other hand, in case of a bond exchange from configuration \textcircled{1} to \textcircled{3} or \textcircled{2} to \textcircled{3}, the chains can no longer bear the axial loads and this exchange will lead to a growth in the voids in \textcircled{3}. In the bulk, such reactions will lead to a reduction in the load-bearing ability and a stretch along  the loading axis. The reverse of this exchange will heal the gap between the two S$-$S chains. This configuration switch (from \textcircled{3} to \textcircled{1} or \textcircled{2}), however, can become increasingly less likely as void growth progresses under the application of stress. Based on the alignment of two S$-$S bonds with respect to the loading axis, configuration \textcircled{3} will result in the incremental stretching of the box, which is the responsible factor towards high secondary creep in vitrimers. 

\begin{figure}[h]
    \centering
    \includegraphics[trim={0cm 0.0cm 2.5cm 0cm},clip=true ,width=0.95\textwidth]{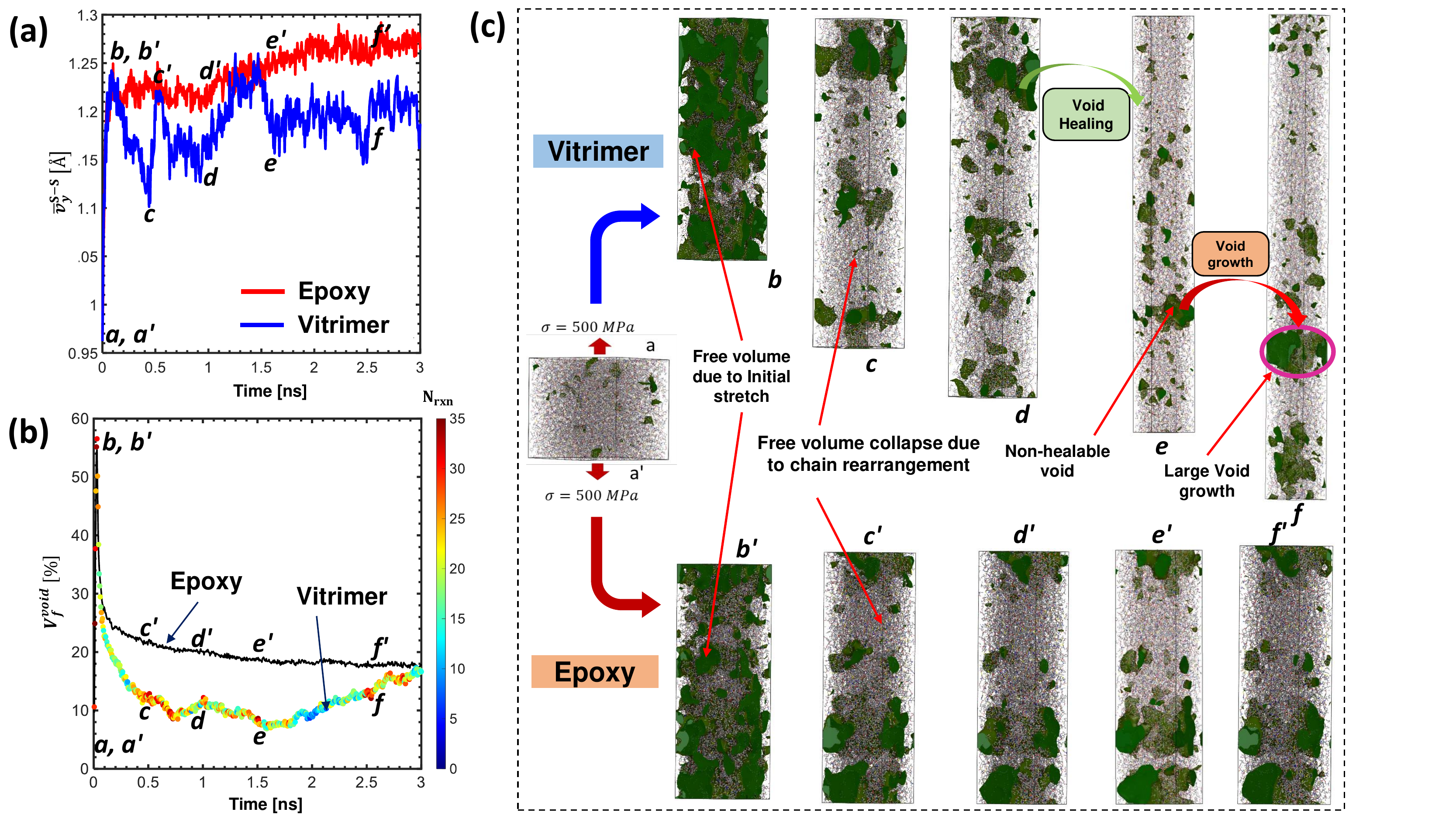} 
    \caption{Comparison of creep in vitrimer and epoxy: (a) mean value of bond vector projection on loading axis (b) evolution of free volume fraction and number of S$-$S reactions in vitrimer, and (c) snapshots of the simulations box with free volume (green region) at different time instances of creep}
    \label{fig:RxNoRxn}
\end{figure}

To evaluate the possibilities of these transformations quantitatively, we look into the values of the S$-$S bond orientation with respect to the loading axis and their evolution over time during creep. Note that the bonds are aligned normal to the loading axis in the void growth configurations \textcircled{3} while in all other configurations, there is a significant component of the bond vector oriented along the loading. The average value of the bond vector projected onto the loading axis is used to quantify the differences in these configurations. A high value indicates that most of the chains are aligned to the loading axis and a relatively lower value indicates the prevalence of type \textcircled{3} configuration which leads to a decrease in stiffness along the loading axis and hence, an increase in creep strain. Figure \ref{fig:RxNoRxn}(a) shows the time variation of the mean value of the S$-$S bond projection on the loading axis ($\overline{v}^\text{S-S}_y$, from Eqn. \ref{meanb_Vector}). In the case of epoxy where these reactions do not occur and any realignment is due only to the chain mobility. This differentiates the projected bond length magnitude from that due to chain mobility versus those occurring due to dynamic reactions.

In Figure \ref{fig:RxNoRxn}(a), the mean bond orientations for vitrimers with and without dynamic bond reactions increase together in the initial phases of the primary creep regime (upto time \emph{b}($\emph{b'}$)). The vitrimer follows the epoxy initially because the total number of dynamic reactions occurring during the initial window is limited due to the small time window and rapid stretch due to high applied stress. The time period from $\emph{b}$($\emph{b'}$) to $\emph{c}$($\emph{c'}$) shows a clear difference between epoxies and vitrimers, at which point the effect of dynamic bonds show up. In epoxies, this region has a relatively stable state of the mean bond orientation with time concomitant with a decrease in free volume occurring due to chain mobility and accommodation. However, in the case of vitrimers, the mean bond projection decreases as the dynamic bond reactions result in the S$-$S bonds aligning towards a plane transverse to the loading axis. Figure  \ref{fig:RxNoRxn}(b) shows the void fraction (in percentage) as a function of time for both epoxy and the vitrimer cases. Initially, both epoxy and vitrimer show a rapid rise in void fraction ($\emph{a}(\emph{a'})$ to $\emph{b}(\emph{b'})$) which is attributed to the observation that the simulation box is not able to instantaneously relax lateral dimensions (perpendicular to loading direction) in response to high applied stress. In the region $\emph{b}(\emph{b'})$) to $\emph{c}(\emph{c'})$, the vitrimer case shows a steeper drop in the void fraction and lower void fraction indicating void healing due to dynamic bonding. 

In vitrimers, the realignment of bonds reaches its peak at time $c$ (Figure  \ref{fig:RxNoRxn}(b)). The time period between $c$ and $e$ is categorized as the secondary creep stage in the vitrimer. In epoxy, the void distribution achieves a stable state around time \emph{c'} and slow changes in free volume beyond \emph{c'} (Figure \ref{fig:RxNoRxn}(c) - bottom) are driven by limited chain mobility. In vitrimers, this regime is related to the formation of smaller voids throughout the volume as shown in Figure  \ref{fig:RxNoRxn}(c) at times \emph{d'} and \emph{e'}. 

The mean orientation increases in this regime as compared to time \emph{c} indicating that the chains are preferring to align along the loading axis, while the void fraction remains somewhat constant (similar to the epoxy). In spite of this, large increases in strain are seen in the secondary creep stage in vitrimers. This can be explained using two processes that are acting in concert. First, there is elongation due to chain rearrangement (similar to epoxy), and second, there is a sudden burst in dynamic bond reactions driving the realignment of S$-$S bonds orthogonal to the loading direction (as seen in the rise and drop in projected bond length between $c$ to $d$ in Figure \ref{fig:RxNoRxn}(a)), which is decreasing the stiffness along the loading axis. This, in turn, further increases the creep strain. In Figure  \ref{fig:RxNoRxn}(b), the points for vitrimers are colored according to the computed number of reactions occurring at each time step. The reactions are seen to accelerate in the secondary creep regime from $c$ to $d$, which led to a significant increase in the creep strain as seen in Figure  \ref{fig:rxnNorxn_strain}(c). 

\begin{figure}[h]
    \centering
    \includegraphics[trim={0cm 7.25cm 0.cm 0cm},clip=true ,width=0.9\textwidth]{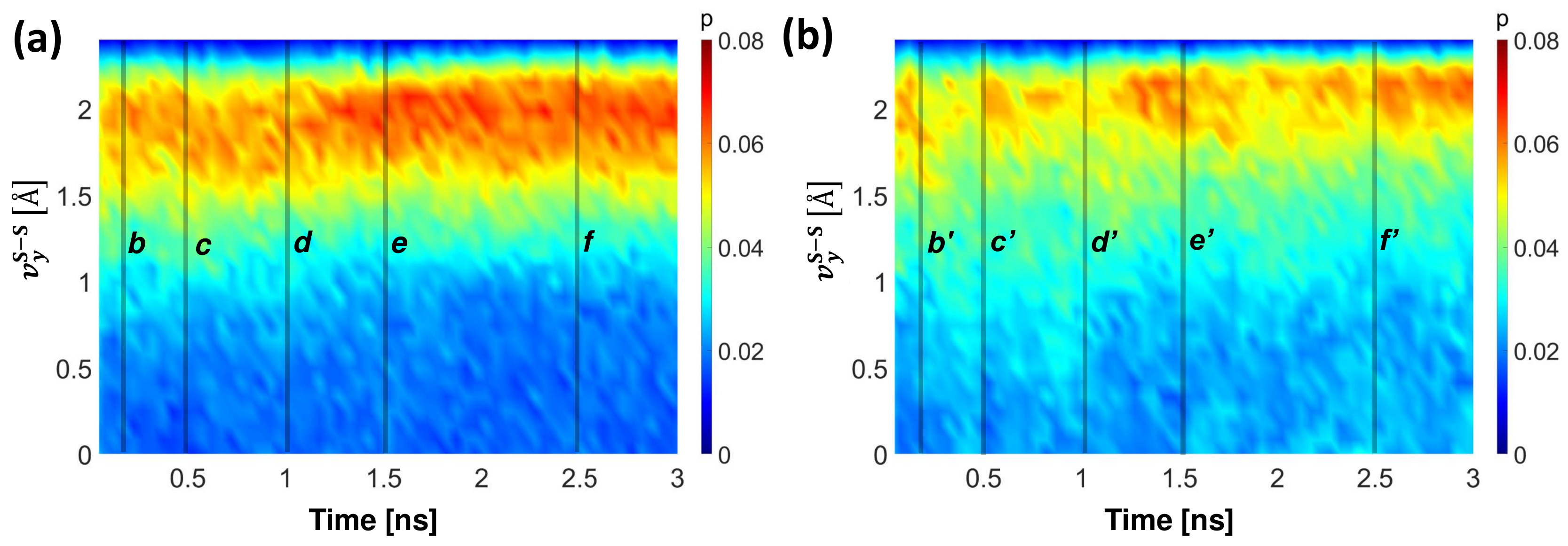}
    \caption{Probability distribution function of S$-$S bond vector projection on loading axis ($v^{S-S}_y$) at each time step for (a) epoxy and (b) vitrimer. The vertical lines mark the key time stamps pertaining to the creep response discussed in the text}
    \label{fig:bondvectorHistogram}
\end{figure}

The stability of void fraction from $c$ to $e$ in vitrimers follows from the fact that the voids created during loading are balanced by the healing of voids via dynamic bonding and subsequent chain rearrangement in new configurations. The effect of dynamic bonding can be more clearly seen from a probability distribution function of the bond projection at each time step. While the plot in Figure \ref{fig:RxNoRxn}(a) only contained the mean at each time step, Figures \ref{fig:bondvectorHistogram}(a) and (b) show time-varying probability density of the bond projection ($v^{S-S}_y$) for epoxy and vitrimer, respectively, showing the complete distribution at each time step. We can see that for epoxy, more number of bonds are clustered towards a bond value of $2.0$ \si{\angstrom} and hence showing a higher mean value with a smaller standard deviation over the $c'$-$e'$ regime. On the other hand, for vitrimer (Figure \ref{fig:bondvectorHistogram}(b)), the probability is less than epoxy near ($v^{S-S}_y=2.0$ \si{\angstrom}) which is the equilibrium bond length of the S$-$S bonds. It also shows a larger spread in the probability throughout the duration of creep. Therefore, the bond exchange reactions lead to more transversely oriented S$-$S bond chains and can result in a configuration \textcircled{2} and \textcircled{3}. The former will contribute towards healing of the voids whenever they appear and the latter will result in a global strain increment as well as void growth. 

\begin{figure}[h]
    \centering
    \centering
    \includegraphics[width=0.6\textwidth]{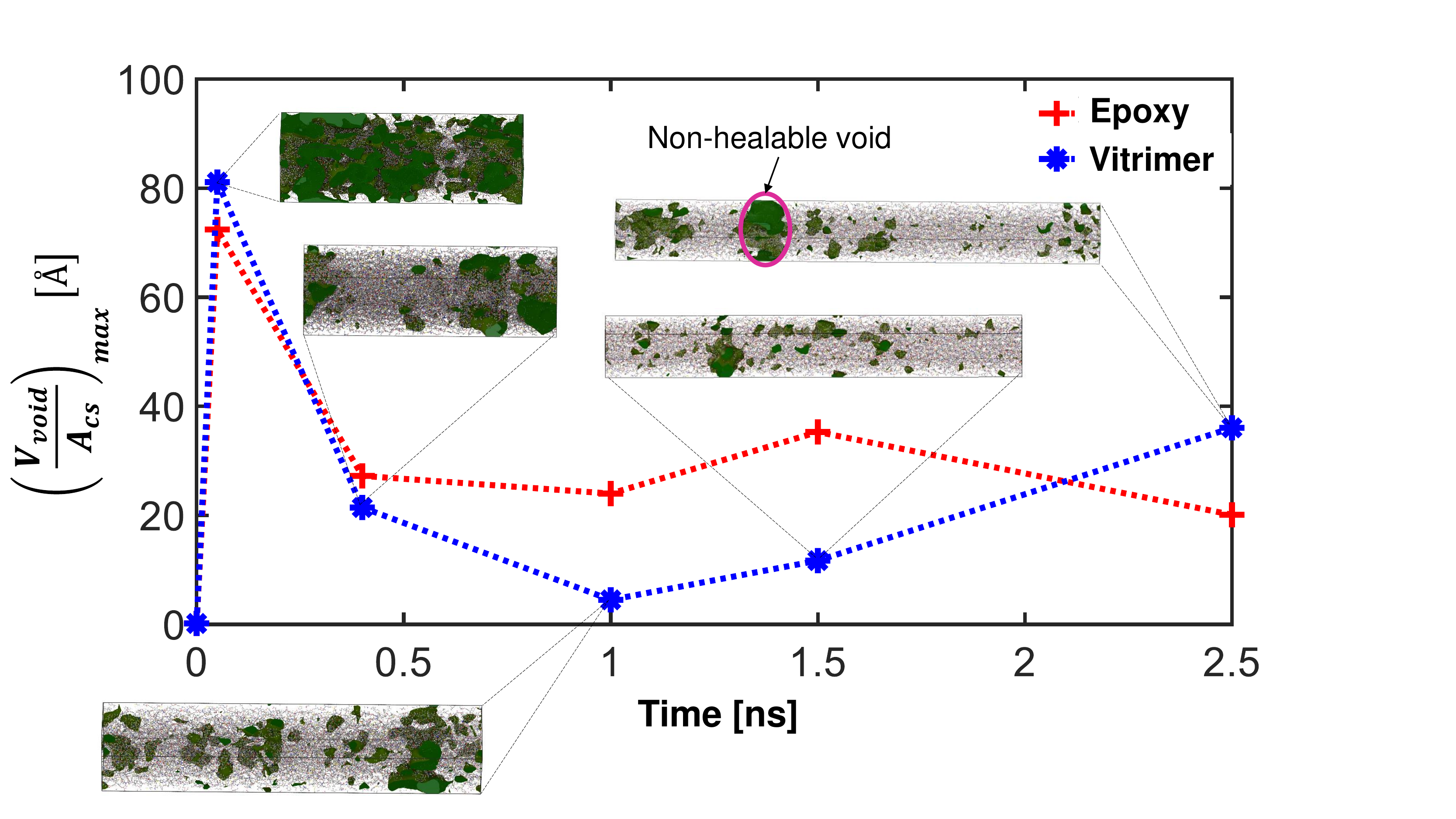}
    \caption{Largest void volume (normalized with the cross-sectional area) at various time stamps during creep of a vitrimer and epoxy. The plotted points correspond to time instances from \emph{a} to \emph{f}.}
    \label{fig:void_size}
\end{figure}

In vitrimers, the region between \textit{e} and \textit{f} is characterized by an increase in the volume fraction of voids. Such behavior is typically associated with tertiary creep. The increase in void-fraction in the vitrimer is related to one large void shown in Figure  \ref{fig:RxNoRxn}(c) that begins to grow and cannot be healed anymore. The number of dynamic bond exchange reactions also declines in this regime and the dynamic bond alignment becomes stable. However, the large void grows resulting in an increase in the creep strain. Figure  \ref{fig:void_size} plots the volume of the largest void in the unit cell normalized to the cross-sectional area. The \textit{y}--axis corresponds to the effective height of the void if the largest void were to span the entire cross-section. An increase in the largest void (non-healable) is seen from $e$ to $f$ in the case of the vitrimer which is shown in the inset and is indicative of tertiary creep in the vitrimer system. \\

The values of stress and reaction probability dictate the extent and the rate of creep in materials. In the previous section, the accelerated creep was studied under extreme conditions. In the following section, we will look into the phenomena under smaller stresses and dynamic bond reaction rates to understand how the creep rates scale in typical laboratory conditions. 

\subsection{Influence of loading}
\begin{figure}[h]
    \centering
    \includegraphics[trim={0cm 0cm 13.5cm 0.0cm},clip=true , width=0.8\textwidth]{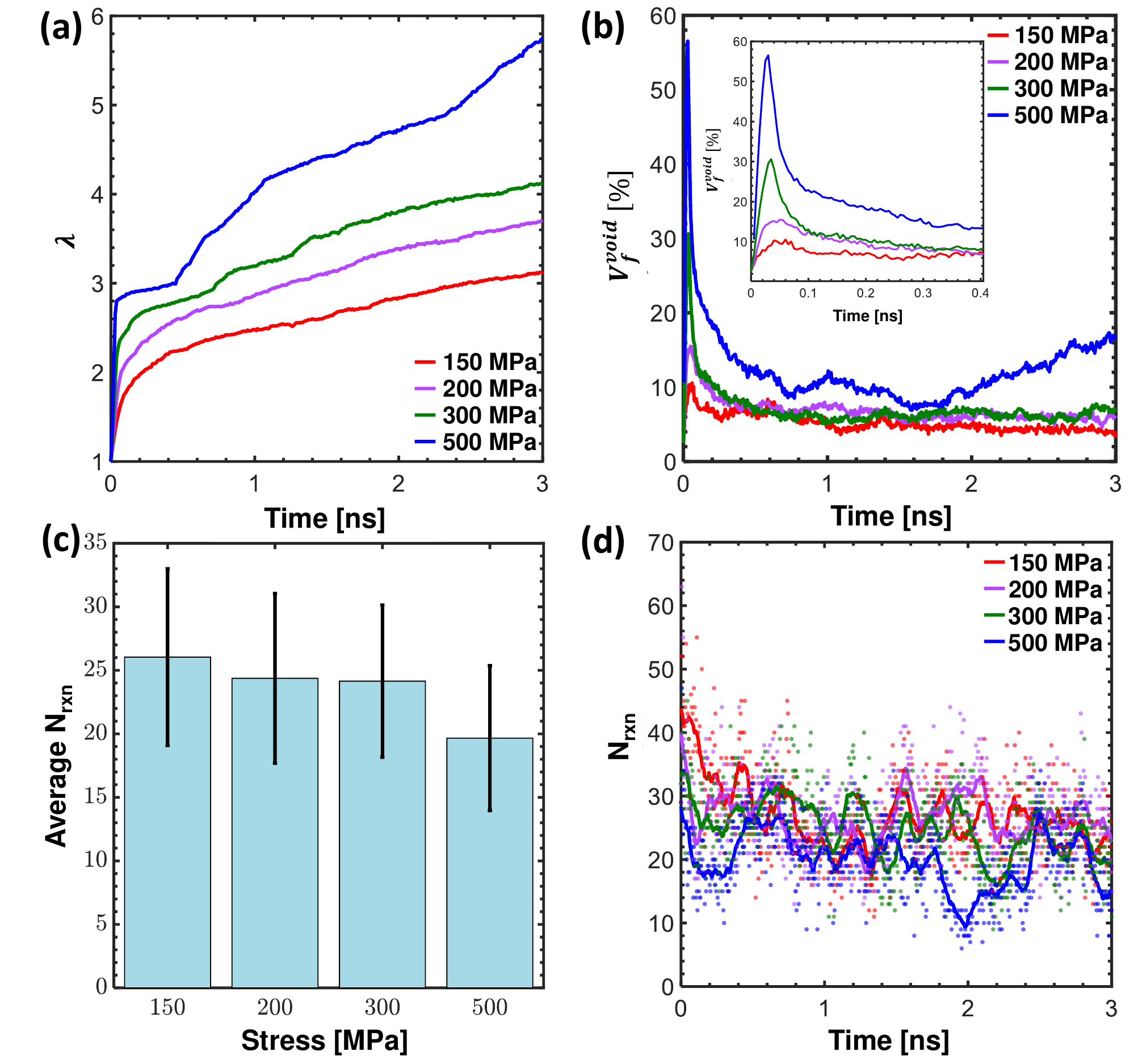}
    \caption{Vitrimer creep under various levels of applied stress (a) stretch ratio ($\lambda$) vs. time and (b) void volume fraction (c) the average number of reactions per loading step, and (d) number of reactions ($\text{N}_\text{rxn}$) at each time step shown in with data points and a moving average shown with line plot}
    \label{fig:load}
\end{figure}

Figure  \ref{fig:load}(a) shows the creep stretch response over time under the application of different values of applied uniaxial tensile stress. It is evident that primary creep strain increases with the magnitude of applied stress. The void fraction shown in Figure \ref{fig:load}(b) remains stable for the lower stress cases over time due to healing processes.  This indicates that the increase in creep strain  during secondary creep is primarily a result of dynamic bonding driving the realignment of bonds orthogonal to the loading direction leading to decreased axial stiffness. The void fraction in the extreme case of 500 MPa is higher initially due to the higher chain mobility, however, healing processes eventually lower the volume fraction of voids to around 10\% beyond which tertiary creep processes take over that increases the void volume fraction again. Tertiary creep behavior is not seen at lower stress levels due to lower chain mobility slowing down the progression of creep as well as a greater probability of healing as evident by the larger number of dynamics reactions at lower applied stress as shown in Figure \ref{fig:load}(c).\\

At higher stresses, the creep strains are higher which increases the total volume of the system and thereby reduces the total number of dynamic S$-$S bonds available to interact per unit volume. This reduction in the number of reactions is seen over time in all the systems as the creep strain increases and is shown in Figure \ref{fig:load}(d). The changes in the number of chemical reactions as a function of time shown in this figure also depict intermittent increases in dynamic reactions accompanying an increase in creep strain. This is followed by intermittent drops in the number of reactions that stabilize the creep strain but at the same time increase the void volume fraction. The creep strain occurs in step-like manner for higher loads due to increases in the types of bond-exchange reactions that lead to void growth and stretch (transformations to configuration \textcircled{3}) followed by void healing processes (transformations to configuration \textcircled{2}). While we observe these step-like increments in creep strain due to the nanoscopic nature of MD simulations (limited number of atoms), a more continuous response is expected to be observed at the macroscopic level, where multiple such step-like increments may be happening at random locations at different times resulting in continuous macroscopic strain response.  For a lower stress value, the stress is not enough to cause such drastic `sudden' increases in creep strain as the chains are unable to rapidly overcome inter-chain interactions (due to lower mobility).

\subsection{Influence of reaction probability}
\begin{figure}[h]
    \centering
     \includegraphics[trim={0cm 0.0cm 13.5cm 0.0cm},clip=true , width=0.8\textwidth]{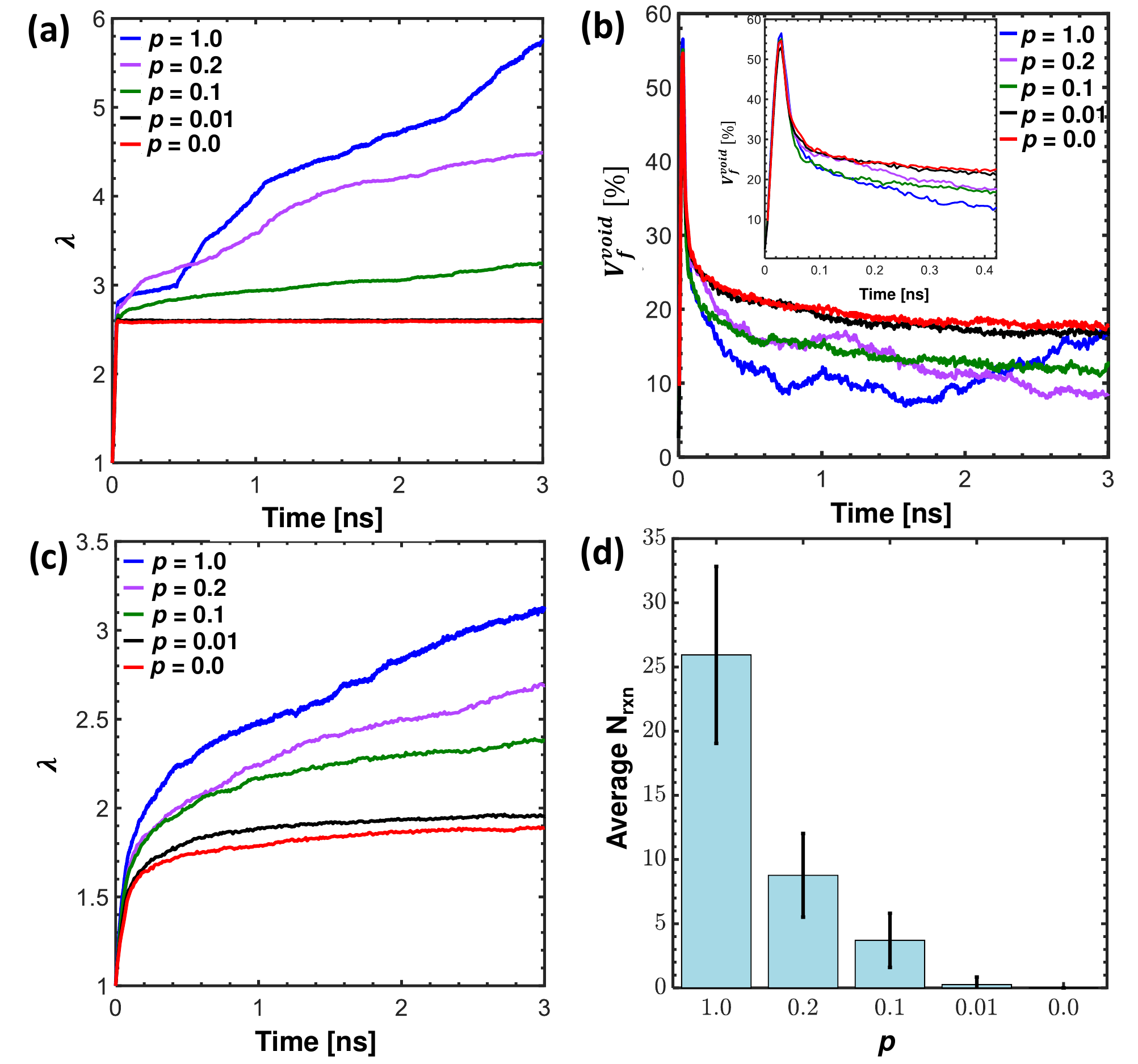}
    \caption{Vitrimer under different S$-$S reaction probabilities at 600 K  (a) stretch ratio vs. time at 500 MPa, (b) Void evolution at 500 MPa (c) stretch ratio vs. time for a lower stress level $\sigma=150$ MPa, (d) average number of reactions per deformation step at 150 MPa}
    \label{fig:prob}
\end{figure}

Interestingly, the evolution of strain in secondary creep in Figure \ref{fig:load}(a) shows that the strain rate remains stable with an increase in applied stress from 150 to 300 MPa. In all these cases, the reaction probability is taken as one, indicating the complete conversion of dynamic bonds when the S-S atoms interact during the simulation. The strain rate is constant for these stress levels indicating that the dynamic bond reaction rates are the controlling factor for differences in the strain rate in secondary creep. A recent experimental study by Hubbard et al. noted that at low temperatures and at low catalyst concentrations, vitrimers simply behave as a traditional epoxy material and the secondary creep rates increase with the amount of catalyst\cite{hubbard2022creep}. This effect of the extent of reactions can be simulated by controlling the probability of the disulfide reactions in our model. Figure \ref{fig:prob} shows the influence of the probability of the dynamic cross-linking reactions for the vitrimer on the creep behavior at an applied stress of 500 MPa. A small increase in primary creep with an increase in the reaction probability is observed. The primary creep strain is relatively unaffected by the reaction probability as the initial strain happens rapidly enough such that the number of dynamic reactions during this stage is low.

However, the reaction probability (as a proxy for the amount of catalyst and temperature) has a strong effect on the strain rate during secondary creep as observed in experiments\cite{hubbard2022creep}. At a very low value of reaction probability (1\%), secondary creep behavior mirrors that of epoxy (Figure  \ref{fig:prob}(a)). However, increasing the reaction probability to $10\%$ causes a significant increase in the creep strain as the finite number of reactions drives more compliant behavior. Reaction probability also has a significant effect on the void fraction evolution as shown in Figure  \ref{fig:prob}(b). Due to void healing in the presence of dynamic reactions, the general trend is that the void fraction is lower with an increase in reaction probability. In the extreme case of a reaction probability of 1.0, the tertiary creep behavior emerges resulting in an increase in the void fraction at later times. Experimental data also show that the tertiary behavior at later times is specific to cases with high amounts of catalysts\cite{hubbard2022creep}. 

The inset of Figure  \ref{fig:prob}(b) shows the rapid increase in void fraction during primary creep is almost identical for all reaction probabilities initially due to the rapidity of loading. Beyond the peak in void fraction, all cases show a drop in the void fraction as the chain mobility leads to a decrease in free volume. However, the general trend of lower void fraction with increased reaction probability is seen during secondary creep once sufficient time is available for the healing of voids via dynamic reactions. Figure  \ref{fig:prob}(c) shows the trends in the creep strain versus time behavior under loading at a relatively lower stress level of 150 MPa. The primary creep regime is more gradual for a lower stress case for all probability cases indicating a closer representation of the laboratory setting of creep experiments (even though the temperature is still elevated).  Figure  \ref{fig:prob}(d) shows that the increase in the number of reactions per deformation step is consistent with the increase in reaction probability.  In Figure  \ref{fig:prob}(c), the trend of the increase in the secondary creep rate with an increase in reaction probability can be seen.  However, the creep strains at 150 MPa in Figure  \ref{fig:prob}(c) are significantly lower than those at 500 MPa in Figure  \ref{fig:prob}(a), and the transition from primary creep to secondary creep is smoother due to lower chain mobility at lower applied stresses. 

\section{Conclusions}
Vitrimers represent the next generation of epoxies, bringing forth the benefits of processability and recyclability. However, the added benefit of dynamic bonds that allow processability also leads to undesired creep behavior. This paper employed MD simulations to understand the nature of creep mechanisms in polymers with dynamic bonding as compared to cases where there is no dynamic bonding. A model disulfide bond exchange-based vitrimer system was investigated using all-atom large-scale molecular dynamics simulations. A novel approach was developed to simulate creep in vitrimers using topology-based bonding and a combination of NVE and NPT simulations to provide stable simulations in the presence of chemical changes. The following conclusions can be drawn from this study:
\begin{itemize}
    \item Vitrimer without reaction (equivalent to epoxy) shows a primary creep response which is driven by chain rearrangement and a very slow secondary creep response due to chain mobility around free-volume. 
    \item Vitrimer with dynamic reactions shows either two or three creep stages depending upon the probability of chemical reactions (equivalent to catalyst concentration and/or temperature). 
    \item In all cases, the first stage is primary creep showing an initial increase in voids due to chain rearrangement as soon as the load is applied followed by a rapid decrease and stabilization of void volume. This behavior is equivalent to epoxy although the void volume is lower in vitrimers due to some healing.
    \item The second stage of creep is the secondary creep where two processes are in concert: elongation due to chain rearrangement (as in primary creep) but also sudden bursts in chemical reactions driving realignment of bonds orthogonal to the loading direction, thus decreasing the stiffness along the loading axis and increasing the creep strain. Very little void growth is seen, as the voids created during loading are balanced by healing via dynamic bonding.
    \item Tertiary creep is seen in high reaction probability and high applied stress cases. In addition to the secondary creep mechanisms of void healing, the growth of an isolated large void(s) occurs in this regime which cannot be healed anymore. 
\end{itemize}
Since the slow secondary creep stage is the rate-determining step in the eventual failure of the material, methodologies to mitigate that should be prioritized in vitrimers. As observed in the simulations, the difference between the secondary creep phenomena in epoxies and vitrimers is the ability of dynamic bonding to reorient the bonds with respect to the loading direction. Thus, chemistry changes or additives that can prevent the initial realignment of dynamic bonds can be an effective strategy to mitigate creep in vitrimers. Indeed, a recent work shows that the addition of metal complexes that decrease chain-to-chain interactions can significantly reduce creep in vitrimers\cite{wang2020facile}. 

\section*{Acknowledgement}
This research was supported by a Seeding To Accelerate Research Themes (START) award titled `Modeling and Characterizing Vitrimer and Vitrimer Composites for Structural Applications' at the University of Michigan. The authors would also like to acknowledge the computational resources and services provided by Advanced Research Computing at the University of Michigan, Ann Arbor.

\section*{Supplementary Information}
Additional data referred to in this paper are included in a supplementary file.

\noindent \textit{Data Availability Statement:} The data that support the findings of this study are available from the corresponding author upon reasonable request.

\bibliographystyle{unsrt}
\bibliography{reference_list}


\section*{Supplementary material}

# SUPPLEMENTARY INFORMATION: UNDERSTANDING CREEP IN VITRIMERS: INSIGHTS FROM MOLECULAR DYNAMICS SIMULATIONS

A PREPRINT

**Gurmeet Singh, Veera Sundararaghavan**[*]
Department of Aerospace Engineering
University of Michigan
Ann Arbor, MI, USA.
gmsingh@umich.edu
veeras@umich.edu

**Vikas Varshney**
Materials and Manufacturing Directorate
Air Force Research Laboratory
Wright-Patterson Air Force Base, OH, USA.
vikas.varshney.2@us.af.mil

April 11, 2023

## Preparation of polymer system for MD simulations

Initially the individual structures of 4-aminophenyl disulfide (AFD) and diglycidyl ether of bisphenol A (DGEBA) are created in Material Studio (shown in Figure S.1). Note that we have created DGEBA structure with epoxied ring open to having a simpler one-step reaction for curing. The epoxy bond in the DGEBA unit is considered to be open as it will reduce the steps in the curing amine reactions. The templates for both primary and secondary amines are created and supplied to the *fix bond/react* command in LAMMPS to enable the curing reactions. Our previous work has detailed information about the curing pre– and post–reaction templates and the curing process followed by annealing of the obtained structure and can referred from [3].

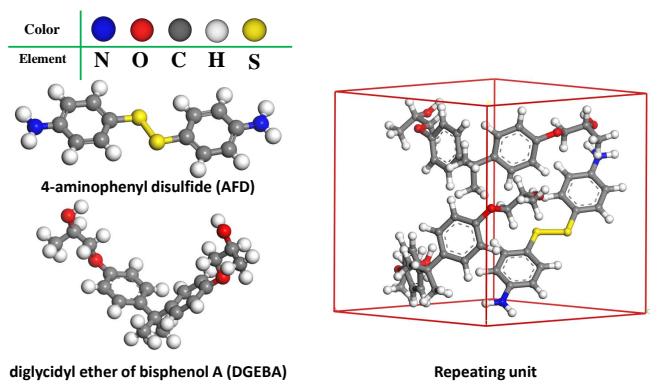

Figure S.1: (a) Monomer structures (left) and the repeat unit box that contains DGEBA:AFD in 2:1 ratio

---

[*]Corresponding author: Prof. Sundararaghavan, Email: veeras@umich.edu, Tel: 734-615-7242

## Simulating creep in vitrimers

The dynamic bond exchange reaction (DBER) of this vitrimer system involves the rearrangement of the chains due to the disulfide bond which is simulated by topology change-based reactions[1, 2]. The pre- and post-reaction templates for this DBER are shown in Figure S.2. The reaction is initiated based on the cutoff distance between the initiator atoms (either one of the green or yellow sulfur atoms).

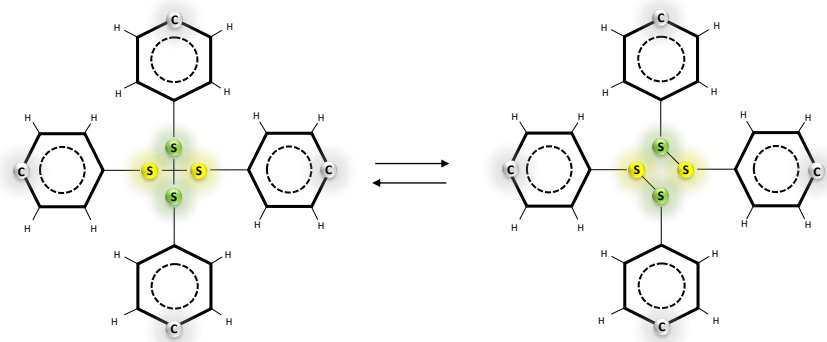

Figure S.2: Reaction template for S-S bond exchange reaction of the vitrimer with pre (on left) and post (on right) reaction templates (each contains 44 atoms). One green and one yellow colored sulfur are the initiator or bonding atoms, and the C atoms shown here denote the edge atom in the templates

To simulate, it is required to prescribe a stress-controlled loading, and as result, there can be sudden large dimension changes in the simulation box. The presence of dynamic bond exchange reaction (local topology change) along with the sudden large change in the box dimensions lead to missing image flags of bonds and atoms across the periodic boundary. Therefore, we use a step-wise alternating loading with NPT (constant number of particles, pressure tensor, and temperature) and reactions under NVE (constant number of particles, volume, and energy) conditions. Figure S.3 shows the effect of the addition of the NVE step along with NPT for an 'epoxy' (vitrimer system with no reactions). This shows that the addition of an NVE step does not introduce any alteration to the chain mobility.

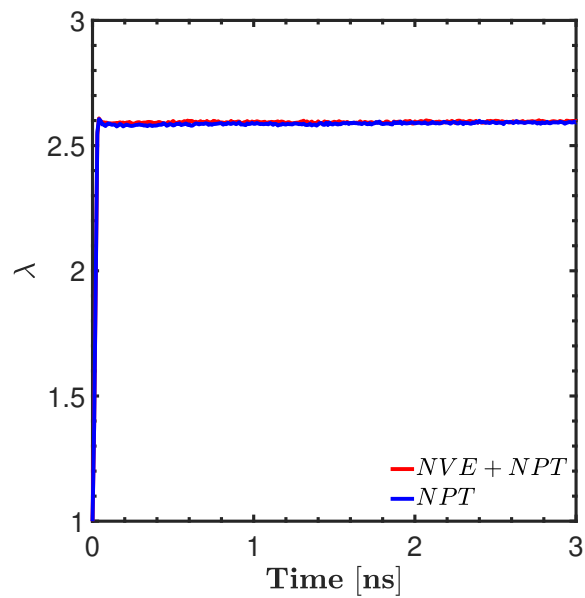

Figure S.3: Comparison of running NPT only and NPT+NVE for the case of vitrimer without reaction


\end{document}